\documentclass[runningheads]{llncs}
\usepackage{graphicx}
\usepackage{booktabs}
\usepackage{vcell}
\usepackage{cite}
\usepackage{amsmath,amssymb,amsfonts}
\usepackage{graphicx}
\usepackage{xcolor}
\usepackage{soul}
\usepackage{diagbox}
\usepackage[normalem]{ulem}
\useunder{\uline}{\ul}{}
\usepackage{mathtools}

\usepackage{url}
\usepackage{paralist}
\usepackage{upgreek}
\usepackage{subfigure}
\usepackage[labelfont=bf]{caption}
\usepackage{mathtools}
\usepackage{multirow}
\usepackage{cleveref}
\usepackage{adjustbox}
\usepackage{float} 
\usepackage{paralist}
\usepackage{schemata}

\usepackage{algorithm}
\usepackage[noend]{algpseudocode}

\usepackage{colortbl}

\newcommand{\etal}{~\textit{et~al.}}

\def\GAa#1{\textcolor{cyan}{{}}} 
 
\begin{document}
\title{Comparative Analysis of Data Augmentation \\ for Retinal OCT Biomarker Segmentation}

\author{Markus Unterdechler\thanks{Correspondence to Markus Unterdechler} \and Botond Fazekas \and Guilherme Aresta, \\and Hrvoje Bogunovi\'c}

\institute{Christian Doppler Laboratory for Artificial Intelligence in Retina, Department of Ophthalmology and Optometry, Medical University of Vienna, Austria\\\email{m.unterdechler@gmx.at, hrvoje.bogunovic@meduniwien.ac.at}}

\authorrunning{M. Unterdechler et al.}

\titlerunning{Data Augmentation for Retinal OCT Segmentation}

\maketitle              
\begin{abstract}
Data augmentation plays a crucial role in addressing the challenge of limited expert-annotated datasets in deep learning applications for retinal Optical Coherence Tomography (OCT) scans. This work exhaustively investigates the impact of various data augmentation techniques on retinal layer boundary and fluid segmentation. Our results reveal that their effectiveness significantly varies based on the dataset's characteristics and the amount of available labeled data. While the benefits of augmentation are not uniform—being more pronounced in scenarios with scarce data, particularly for transformation-based methods—the findings highlight the necessity of a strategic approach to data augmentation.  It is essential to note that the effectiveness of data augmentation varies significantly depending on the characteristics of the dataset. The findings emphasize the need for a nuanced approach, considering factors like dataset characteristics, the amount of labelled data, and the choice of model architecture.




\keywords{Machine Learning \and  Data Augmentation \and  Retinal OCT \and fluid segmentation}
\end{abstract}

\section{Introduction}

Optical coherence tomography (OCT) is a widely used 3D retinal imaging modality, aiding in the assessment of biomarkers like retinal layer thickness and fluid volume, key for diagnosing and treating retinal diseases such as age-related macular degeneration (AMD), retinal vein occlusion (RVO), and diabetic macular edema (DME), the leading causes of blindness. In particular, accurate detection and quantification of fluid accumulations, such as intraretinal fluid (IRF), subretinal fluid (SRF), and pigment epithelial detachment (PED), are important for clinical decision-making due to their direct relation to disease activity. Automated OCT segmentation algorithms offer potential but their development and validation is limited. 

Data augmentation enriches labeled datasets through a variety of transformations, from rotations and scalings to more advanced techniques such as introducing specialized noise, specific to OCT scans or using Generative Adversarial Networks (GANs) to generate synthetic data \cite{rebuffi2021data, konidaris2019generative, chlap2021review}. This process enhances model resilience against image variation and improves generalization \cite{shorten2019survey}. However, the effectiveness of augmentation varies with data characteristics, necessitating thorough comparative analysis to optimize its application across diverse domains.



In the field of retinal fluid segmentation on OCT scans, researchers use a plethora of methods for data augmentation. However, the absence of a standardised comparative study has resulted in a lack of  understanding of the most effective approach. In this study we evaluate different data augmentation techniques to assess their influence on training performance relative to dataset characteristics. The results will be used to guide the selection of appropriate strategies for specific use cases.

\paragraph{\textbf{Related Work}} Chlap\etal \cite{chlap2021review} categorised data augmentation methods for deep learning-based medical imaging applications (Supplement Fig.~S1). Augmentation techniques were categorized into four groups: \emph{Basic}, \emph{Deformation}, \emph{Deep Learning} and \emph{Other}. The Basic transformations include linear affine transformations and any adjustments to pixel intensity levels. Deformation covers various elastic transformations, while the Deep Learning category refers to augmentation techniques that generate synthetic data using deep learning techniques. They found that the basic transformations are the most prevalent due to their simplicity and rapid implementation during training \cite{buslaev2020albumentations}. 
For instance, Bar-David\etal \cite{bar2021impact} showed that shallow networks can outperform deeper networks by using data augmentation in a pixel-wise segmentation task. They used various augmentation techniques, such as affine transformations, flipping, noise induction, contrast saturation, and elastic transformations. 

\begin{figure}[bt]
  \centering
  \includegraphics[width=0.8\textwidth]{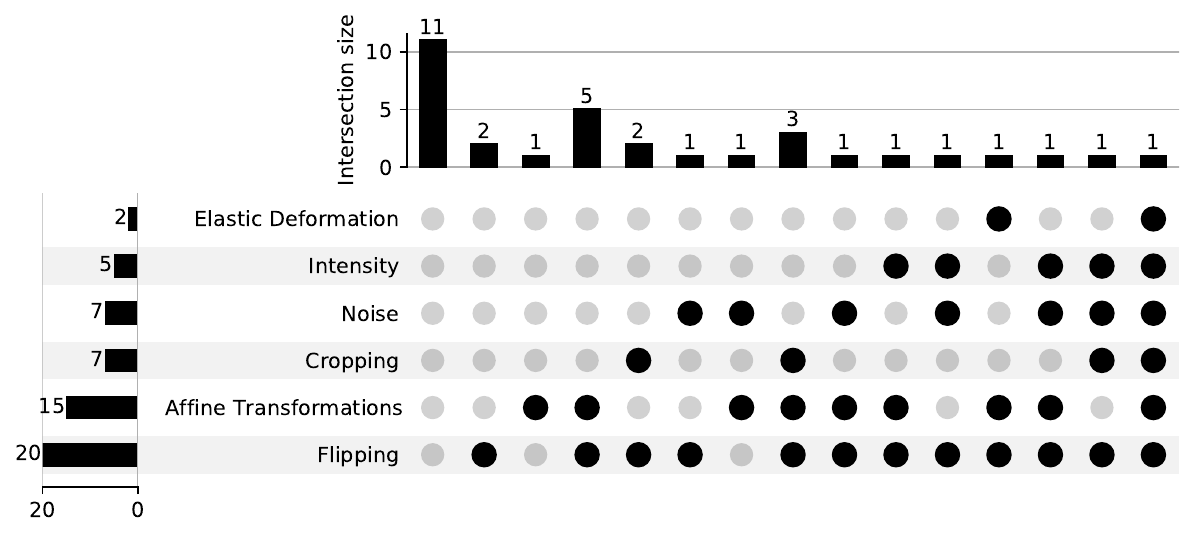}
   \caption{Overview of the number of papers found that use certain data augmentation techniques in retinal OCT segmentation studies.}
  \label{fig:da_oct_literature_review}
  \vspace{-1em}
\end{figure}

Currently, there is no standard data augmentation approach for retinal OCT. Reviewing 33 works on this topic (Fig.~\ref{fig:da_oct_literature_review}), the most common augmentations were methods such as random horizontal flipping and linear affine transformations, due to their computational efficiency and effectiveness in reflecting biological structures (e.g., \cite{he2023longitudinal,pekala2019deep,he2019fully}). These methods have demonstrated their capability in effectively decreasing overfitting \cite{li2019parallel}. Cropping reduces the input size but has a comparable effect to affine translations \cite{shorten2019survey}. Some studies have focused on dataset standardization rather than augmentation \cite{stromer2020correction,ruan2019multi}. Noise-related augmentations, primarily Gaussian noise \cite{ouyang2019accurate,xie2023deep}, but also speckle noise \cite{lazaridis2020bio}, blurring, and sharpening effects, were observed, along with intensity adjustments like brightness, contrast, and HSV color space alterations  \cite{pekala2019deep,li2020deepretina}. Deformation-related augmentations were less common \cite{ouyang2019accurate,kepp2019topology}.

\paragraph{\textbf{Contributions:}} There are significant differences in augmentation choices among studies addressing the same domain and tasks, underscoring the need for a comprehensive assessment to establish a state-of-the-art approach. This foundational study contribute to the field of automated retinal OCT analysis by conducting exhaustive evaluation on the influence of traditional data augmentation techniques on the two clinically most relevant tasks of retinal layer boundary and fluid segmentation. We propose metrics that can be used to describe the intrinsic characteristics of the dataset, and use them to support the hypothesis of when augmentation types should be used. Furthermore, we demonstrate techniques to improve generalization across different scanner types. Thus, we provide \emph{guidelines}, which data augmentation techniques to use, taking into account the objective of the task and the characteristics of the datasets. We focus on traditional augmentation techniques as these techniques are the most widely used in the literature, so a comparative study of these techniques will have the greatest impact in this area.

\section{Materials and Methodology}

\paragraph{\textbf{Datasets}}
The effect of data augmentation on retinal segmentation tasks were evaluated on two publicly available retinal OCT datasets and on a private one.

The public~\emph{MSHC} dataset \cite{he2019fully} comprises of 14 volumes from healthy controls (HCs) and 21 with multiple sclerosis (MS). Each volume has 49 B-scans obtained with a Spectralis (Heidelberg Engineering, Germany) scanner, with 9 manually annotated surfaces.

The public~\emph{RETOUCH} dataset \cite{2019_Bogunovic} comprises of 112 OCT volumes gathered from patients with macular edema due to AMD or RVO. It contains manual annotations for intraretinal fluid (IRF), subretinal fluid (SRF), and pigment epithelial detachment (PED). The training set includes 70 OCT volumes and the test set consists of 42 scans, which were captured in approximately equal numbers using three different devices: Spectralis, and the considerably noisier Cirrus HD-OCT (Zeiss Meditec, Germany) and T-1000/T-2000 (Topcon, Japan).

The private~\emph{Multibiomarker} dataset has 85 macula-centered Spectralis OCT scans from different patients, ranging from 18 to 97 B-scans in resolution. It was obtained during routine clinical check-ups of patients with neovascular AMD patients undergoing treatment, providing insights into real-world clinical situations, with 11 manually annotated layers.

\subsection{Acquisition-related Scan Characterization}
We hypothesise that the impact of different data augmentation techniques depends on acquisition-related characteristics of retinal OCTs, namely \emph{Symmetry}, \emph{Contrast}, \emph{Signal-to-noise-ratio} and \emph{Alignment}. 

\begin{figure}[tb]
  \centering
  \includegraphics[width=0.9\textwidth]{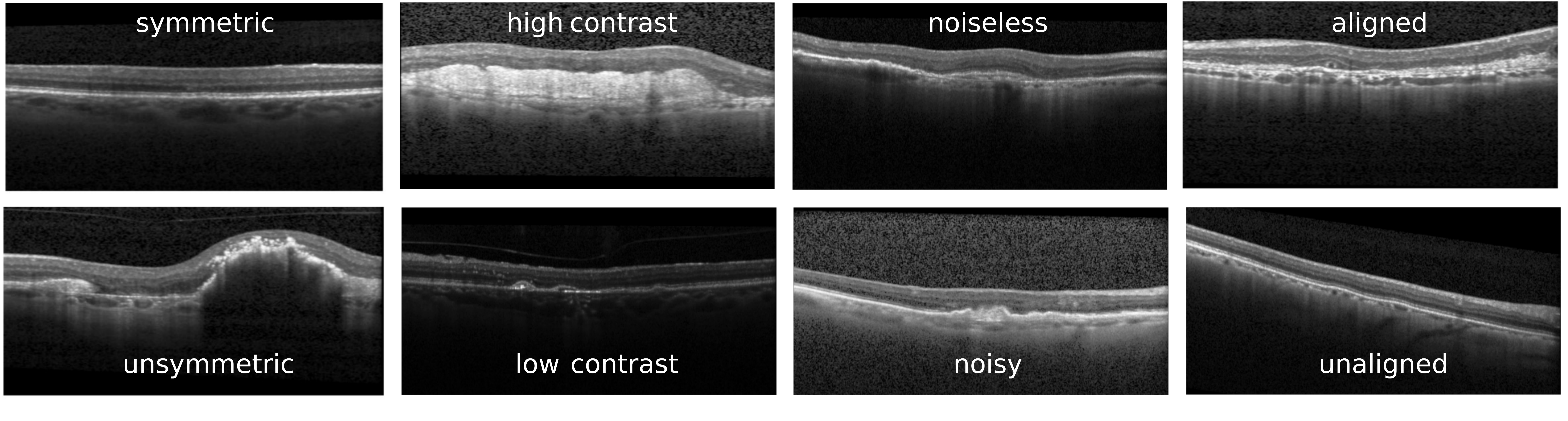}
   \caption{High and low value examples of OCT scans per metric.}
  \label{fig:Metric_Examples}
  \vspace{-2em}
\end{figure}

\emph{Alignment} measures the degree of the horizontal alignment of the retina within OCT scans. It is calculated based on the highest pixel value per column in the scan image, corresponding to the high intensity of the RPE layer. \emph{Symmetry} addresses the retinal structure similarity between the left and right portions of a B-scan. First, the retina is flattened, by rotating the image such the RPE layer becomes horizontally, and then the left and right halves are compared using the Normalised Cross-Correlation Coefficient. To reduce sensitivity to noise, a small Gaussian blur filter is applied. \emph{Contrast} relates to the difference in luminance between different parts of the image, which significantly affects the ability to assess different biomarkers. As a measurement, the standard deviation of the intensity histogram was used. The Signal-to-Noise ratio (SNR)~\cite{johnson2006signal} is used to quantify the \emph{Noise} level in OCT scans. Each metric provides a float value. Figure~\ref{fig:Metric_Examples} illustrates extreme cases.

\subsection{Data Augmentation techniques}
  
We assess data augmentation techniques from five distinct groups: Transformation, Deformation, Intensity, Noise, and Domain Specific (Table~\ref{tab:augmentation_categories}). The evaluated hyperparameters ranges are displayed in Table~S1 in the Supplement.
\begin{table}[htb]
  \centering
  \renewcommand{\arraystretch}{1.0}
  \adjustbox{max width=\textwidth}{
    \setlength{\tabcolsep}{4pt} 
    \small 
    \begin{tabular}{ccccc} 
      \toprule
      \textbf{Transformation}                                                           & \textbf{Deformation}                                                        & \textbf{Intensity}                                                                             & \textbf{Noise}                                                                                    & \textbf{Domain Specific}   \\ 
      \midrule
      \vcell{
        $\left.\begin{tabular}[p]{@{}c@{}}
        Rotation\\
        Shear\\
        Scale\\
        Translate
        \end{tabular} \right\}$ \rotatebox[origin=c]{90}{Affine}
      } &
      \vcell{\begin{tabular}[b]{@{}c@{}}Short Elastic\\Long Elastic\end{tabular}} & \vcell{\begin{tabular}[b]{@{}c@{}}Contrast Adjustment\\Histogram Matching\end{tabular}} & 
      \vcell{\begin{tabular}[b]{@{}c@{}}Gaussian Noise\\Speckle Noise\\SVD Noise Transfer\end{tabular}} & \vcell{Vessel Simulation}  \\[-\rowheight]
      \printcelltop                                                                     & \printcelltop                                                               & \printcelltop                                                                                  & \printcelltop                                                                                     & \printcelltop              \\
      \\
      \bottomrule
    \end{tabular}
  }
  \caption{Overview of the data augmentation categories}
  \label{tab:augmentation_categories}
\end{table}
\vspace{-3em}



\textbf{Transformation}-based techniques, encapsulated under the term \textit{Affine transformations}, introduce geometric variations, e.g., rotation, translation, scaling, and shearing to mimic real-world data fluctuations. This approach ensures model robustness against orientation, position, scale, and shape variations by applying transformations uniformly to images and labels.

\textbf{Deformation} augmentations utilize 2D elastic transformations to introduce distortions to the data. This technique can simulate local non-linear, wave-like deformations that reflect anatomical variability observed in patients. Two settings are explored: \textit{Long Elastic} with wider spacing and \textit{Short Elastic} with finer granular deformations.

\textbf{Intensity}-related augmentation methods, such as \emph{Contrast Adjustment} and \emph{Random Histogram Matching}, change pixel values without altering spatial arrangement. \emph{Contrast Adjustment} changes contrast to delineate light and dark areas more or less distinctly, whereas \emph{Random Histogram Matching (RH Matching)} aligns the pixel value distribution of a target image with that of a randomly chosen source image. This ensures similar contrast and brightness characteristics between the target and source images, to reduce the model's sensitivity to illumination variations.

\textbf{Noise} augmentations add artificial noise to OCT scans, enhancing model robustness by training on data mimicking real-world imperfections. \textit{Gaussian noise} introduces random variations to pixel values based on a Gaussian distribution, adjusting image brightness and texture. \textit{Speckle noise}, characteristic of OCT due to coherent light interference, is simulated through granular pixel fluctuations corresponding to the Poission distribution. \textit{Singular Value Decomposition Noise Transfer (SVD Noise Transfer)} proposed by Koch\etal\cite{koch2022noise} transfers noise characteristics between OCT scans via singular value decomposition, enriching training data with realistic noise patterns.


\textbf{Domain-Specific} augmentation leverages specialized knowledge to tailor data enhancement for specific applications. In retinal OCT scans, the model's robustness against variations caused by vessel shadows can be improved by simulating them during training. \textit{Vessel Simulation} randomly sets the count, dimensions, and placement of simulated shadows within predefined limits and applies corresponding shading to image columns to replicate vessel appearances. This process helps the model learn to recognize and adapt to vessel-related variations in OCT images.

\section{Experiments and Results}


\paragraph{\textbf{Segmentation architecture}} The layer and fluid segmentation was performed using a U-Net-based retinal biomarker segmentation architecture proposed by He\etal \cite{he2021structured}, with the encoder replaced by a ResNet-34.

\paragraph{\textbf{Layer segmentation task}} In the layer segmentation experiments, augmentation types were applied to every scan with varying intensity and evaluated under three scenarios: using all volumes, training with 4 patient volumes containing 7 slices each, and with 2 patient volumes of 7 slices. The training was conducted using a batch size of 4 and lasted for 200 epochs. A learning rate of 1e-3 was utilized, with a warm-up period of 20 epochs and cosine decay. The Adam optimizer was employed for optimization during training. A combination of DiceCE, KL and L1 loss was used as proposed by He\etal. The 5-fold cross-validation outcomes for these tasks on the \emph{MSHC} and \emph{Multibiomarker} datasets are detailed in Fig.~\ref{fig:layer_rmsediff}, with per-layer results in the Supplement (Tables~S2-4 and ~S5-7). The $\Delta$RMSE, representing the signed difference between augmented and baseline results, reveals either an improvement ($\Delta$RMSE $< 0$) or a deterioration ($\Delta$RMSE $> 0$) compared to no augmentation scenarios. The Wilcoxon signed-rank test was used to test statistical significance.



\begin{figure}[bt]
  \centering
  \includegraphics[width=1.0\textwidth]{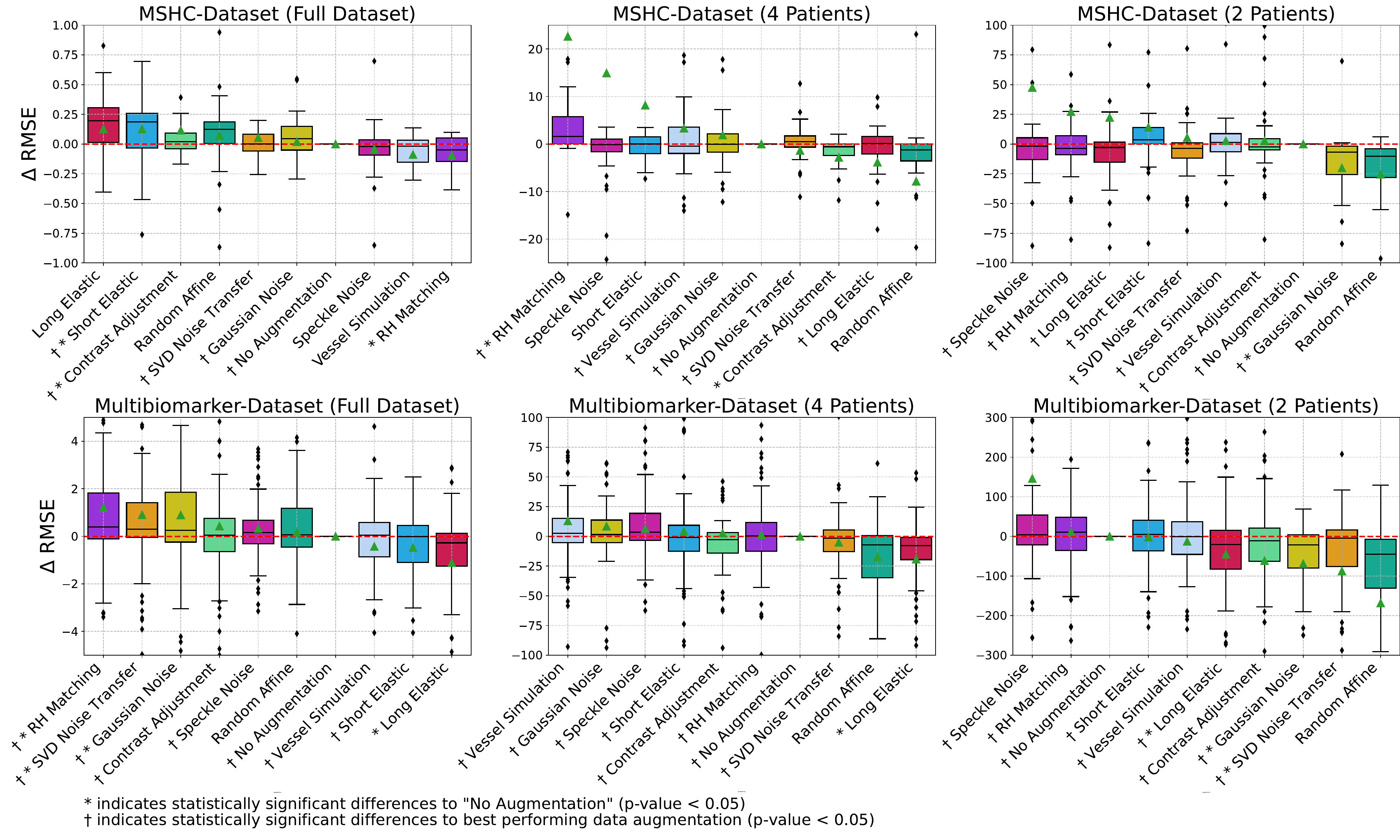}
\caption{RMSE between augmentation types compared to 'No Augmentation' augmentation.}
  \label{fig:layer_rmsediff}
  \vspace{-2em}
\end{figure}

\emph{MSHC}, being more standardized and of higher quality, facilitates easier segmentation of retinal layers compared to the diverse \emph{Multibiomarker}, which exhibits structural abnormalities and noise. This difference in dataset characteristics leads to better segmentation performance on \emph{MSHC}. Deformation-based augmentation adversely affects the structurally intact \emph{MSHC} but benefits the diverse \emph{Multibiomarker} dataset.

The impact of data augmentation on the \emph{MSHC} is minimal with a full training set but grows as the dataset size decreases, with Random Affine Transformations becoming particularly vital for smaller samples, a trend also observed in the \emph{Multibiomarker} dataset. 

The influence of data augmentation, analyzed based on \textit{Symmetry}, \textit{Contrast}, \textit{Noise}, and \textit{Alignment}, is detailed in Fig.~\ref{fig:ms_tdn_RMSE_Metrics} for \emph{MSHC}, with comprehensive results in the Supplement (Fig.~S2).
The \emph{Transformation} and \emph{Deformation}-based augmentations enhances segmentation performance on unsymmetrical and unaligned samples, with a marginal decline observed on symmetric and aligned samples, where Affine Transformations exhibit minimal adverse effects. 

\begin{figure}[bt]
  \centering
  \includegraphics[width=1.0\textwidth]{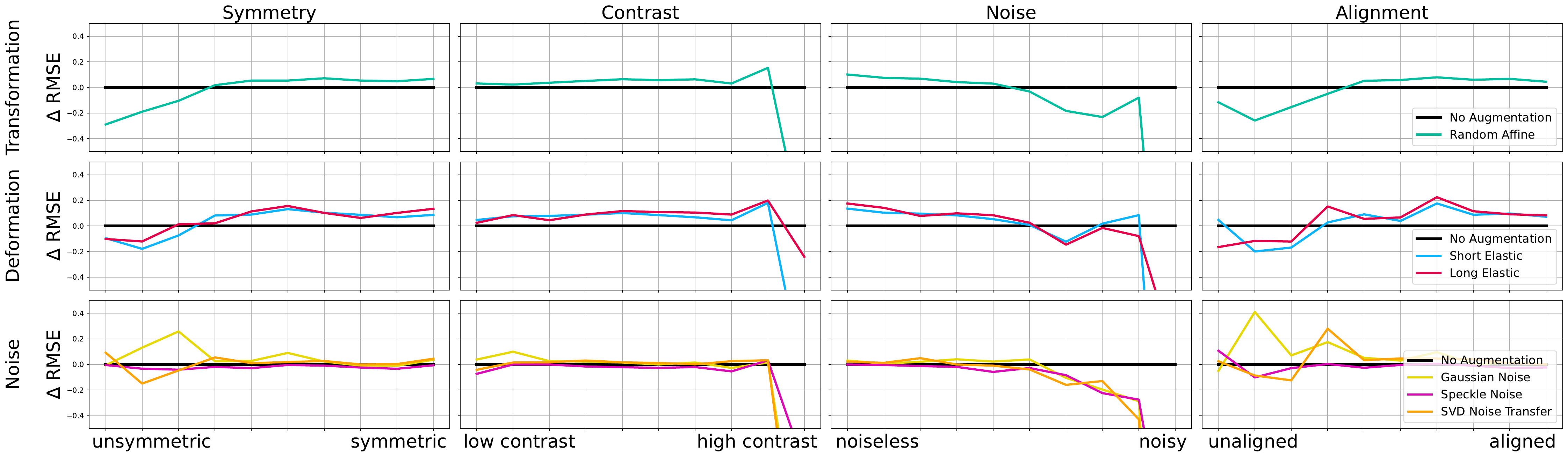}
   \caption{\emph{MSHC}-Dataset Full: RMSE distance with respect to OCT Scan metrics.}
  \label{fig:ms_tdn_RMSE_Metrics}
  \vspace{-1em}
\end{figure}



\begin{figure}[bt]
  \centering
  \includegraphics[width=1.0\textwidth]{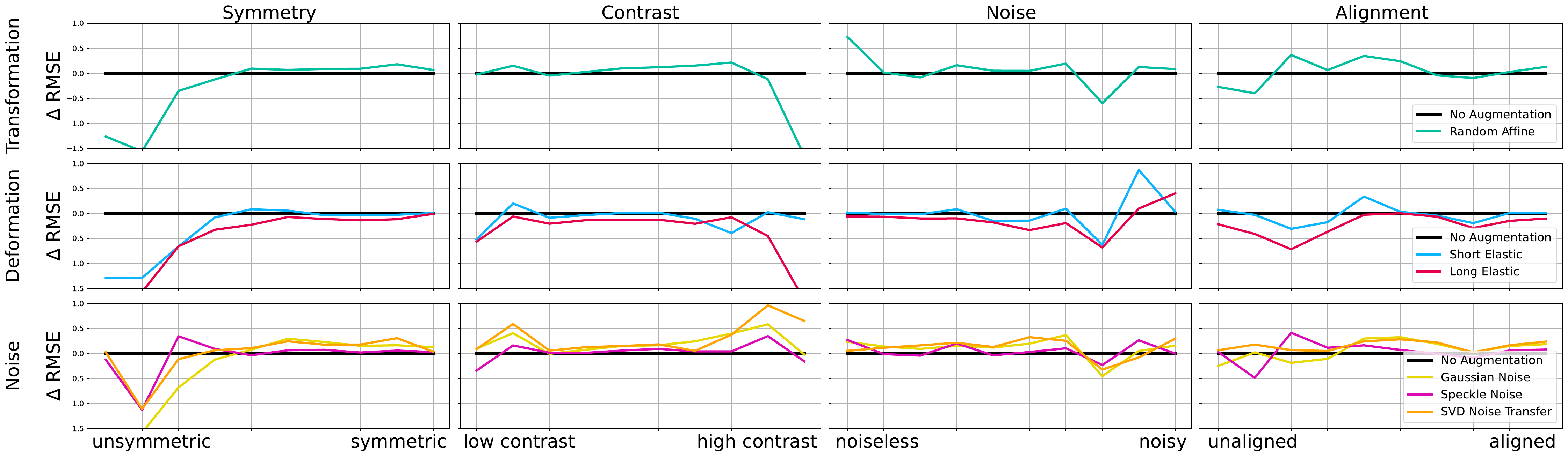}
   \caption{\emph{Multibiomarker}: RMSE distance with respect to OCT Scan metrics.}
  \label{fig:multibiomarker_tdn_RMSE_Metrics}
  \vspace{-1em}
\end{figure}

The analysis of the \emph{Multibiomarker} dataset (Fig.~\ref{fig:multibiomarker_tdn_RMSE_Metrics}), reveals different outcomes compared to \emph{MSHC}. Here, \textit{Deformation}-based augmentations not only benefit symmetrical samples but also demonstrate a slight drawback on straight retinal structures, emphasizing their advantage for structural irregularities. In contrast, \textit{Noise}-related augmentations do not enhance segmentation for noisier scans within this dataset, suggesting a universal detrimental effect due to the already prevalent noise in the scans.

\begin{figure}[tb]
  \centering
  \includegraphics[width=1.0\textwidth]{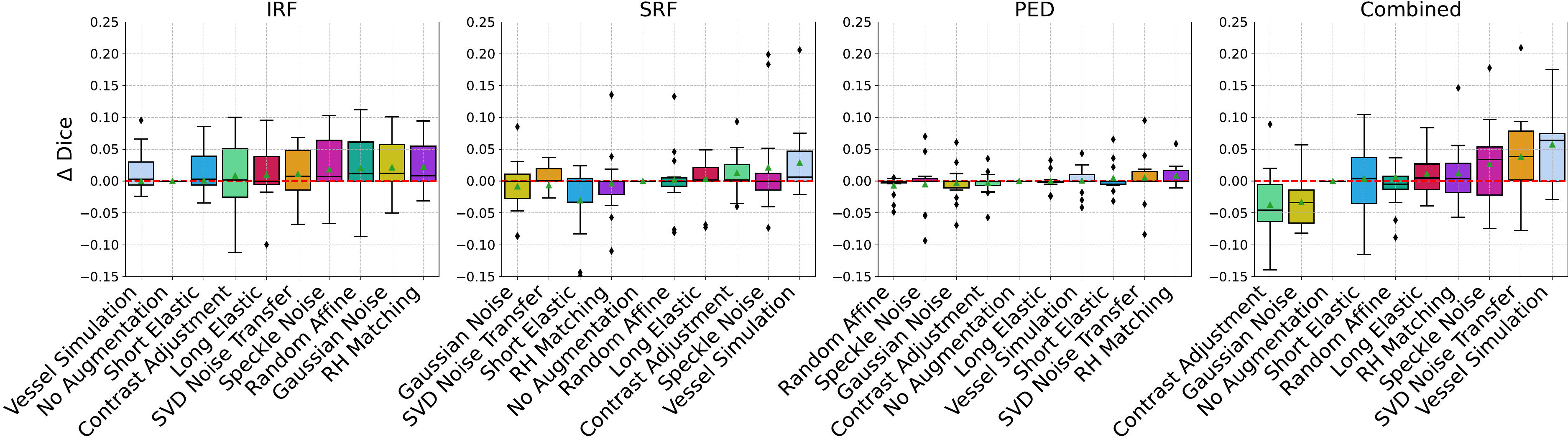}
\caption{Dice Score difference between augmentation types compared to 'No Augmentation' for fluid segmentation on Spectralis scans.}
  \label{fig:dicediff_retouch_spectralis}
  \vspace{-1em}
\end{figure}

\paragraph{\textbf{Fluid segmentation task}} In fluid segmentation experiments trained on the \emph{RETOUCH} using Spectralis scans, most augmentation methods enhanced the segmentation performance (Fig.~\ref{fig:dicediff_retouch_spectralis}). The training lasted for 600 epochs and used a batch size of 8, with augmentations applied at a probability of 0.33. A learning rate of 1e-3 was applied, along with a 20-epoch warm-up period and cosine decay. The Adam optimizer, paired with the Dice loss, was used for optimization during training. While outcomes generally improved, specific augmentations like \emph{Vessel Simulation}, despite scoring highest overall, had a unique negative impact on IRF segmentation, underscoring the nuanced effect of different augmentations on segmentation accuracy.


\begin{figure}[tb]
  \centering
  \includegraphics[width=1.0\textwidth]{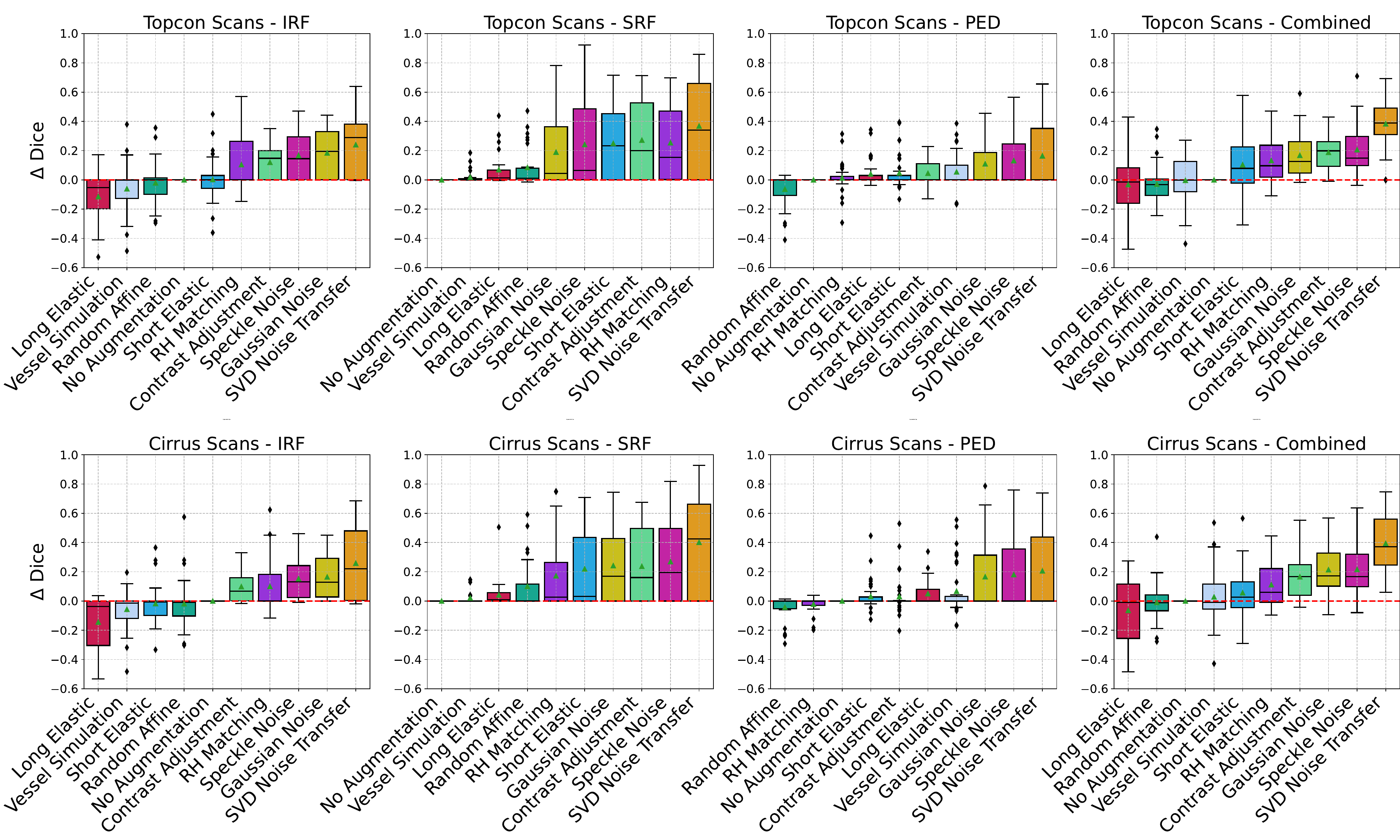}
\caption{Dice Score difference between augmentation types 
for fluid segmentation cross-device \emph{generalization} performance on Topcon- and Cirrus-scans.}
  \label{fig:dicediff_retouch_topcon_cirrus}
  \vspace{-1em}
\end{figure}

The cross-device generalization capability was investigated by evaluating the models, which were trained on Spectralis, on Topcon and Cirrus scans (Fig.~\ref{fig:dicediff_retouch_topcon_cirrus}). In particular, noise-related augmentations such as SVD Noise Adaptation were shown to significantly improve these segmentation results. On the other hand, Affine transformations had a negative impact when evaluated on Cirrus and Topcon scans, an effect not seen in the Spectralis results.

\section{Conclusion}

We comprehensively assessed the effect of data augmentation on retinal biomarker segmentation on OCT scans. The experiments highlight the importance of spatial transformation-based techniques, including affine transformations and elastic deformations on retinal OCT scans. While noise-related augmentations are relatively common in the literature, we found that they can have a strongly detrimental effect on already noisy datasets, while providing benefits only in a few cases. Data augmentation, especially SVD Noise Adaption, has demonstrated its ability to boost generalization across different scanner types. Future research aims to improve the value of augmentation strategies by focusing on adaptive techniques tailored to the dataset characteristics.
\newpage
\bibliographystyle{splncs04}
\bibliography{Paper-0022}

\end{document}


\begin{table}[htbp]
\centering
\caption{Data Augmentation - Hyperparameters}
\label{DA_Hyperparameters}
\small
\renewcommand{\arraystretch}{1.2}
\resizebox{\linewidth}{!}{%
\begin{tabular}{p{6cm}|p{9cm}}
\hline
\textbf{Augmentation} & \textbf{Hyperparameters} \\
\hline
Random Affine & Rotation: from -0.2 to 0.2 \\
& Shear: from -0.2 to 0.2 (both axis) \\
& Translate: from -0.3 to 0.3 (axis 0) \\
& Scale: from -0.3 to 0.3 (axis 0) \\
\hline
Long Elastic & Spacing: 100px horizontal and 100px vertical \\
& Magnitude: from 0.0px to 10.0px \\
\hline
Short Elastic & Spacing: 20px horizontal and 20px vertical \\
& Magnitude: from 0.0px to 2.0px \\
\hline
Contrast Adjustment & $\gamma$: from 0.9 to 1.4 \\
\hline
RH Matching & Ratio: from 0.5 to 1.0 \\
\hline
Gaussian Noise & $\sigma$: from 0.0 to 0.2 \\
\hline
Spckle Noise & $\sigma$: from 0.0 to 0.4 \\
\hline
SVD Noise Transfer & k: 40 \\
& hm\_degree: 0.9 \\
\hline
Vessel Simulation & Nr of vessels: from 0 to 5 \\
& Width: from 10 to 100 \\
& Shading factor: from 0.5 to 0.7 \\
\hline
\end{tabular}
}
\end{table}

\begin{figure}[htb]
  \centering
  \includegraphics[width=0.6\textwidth]{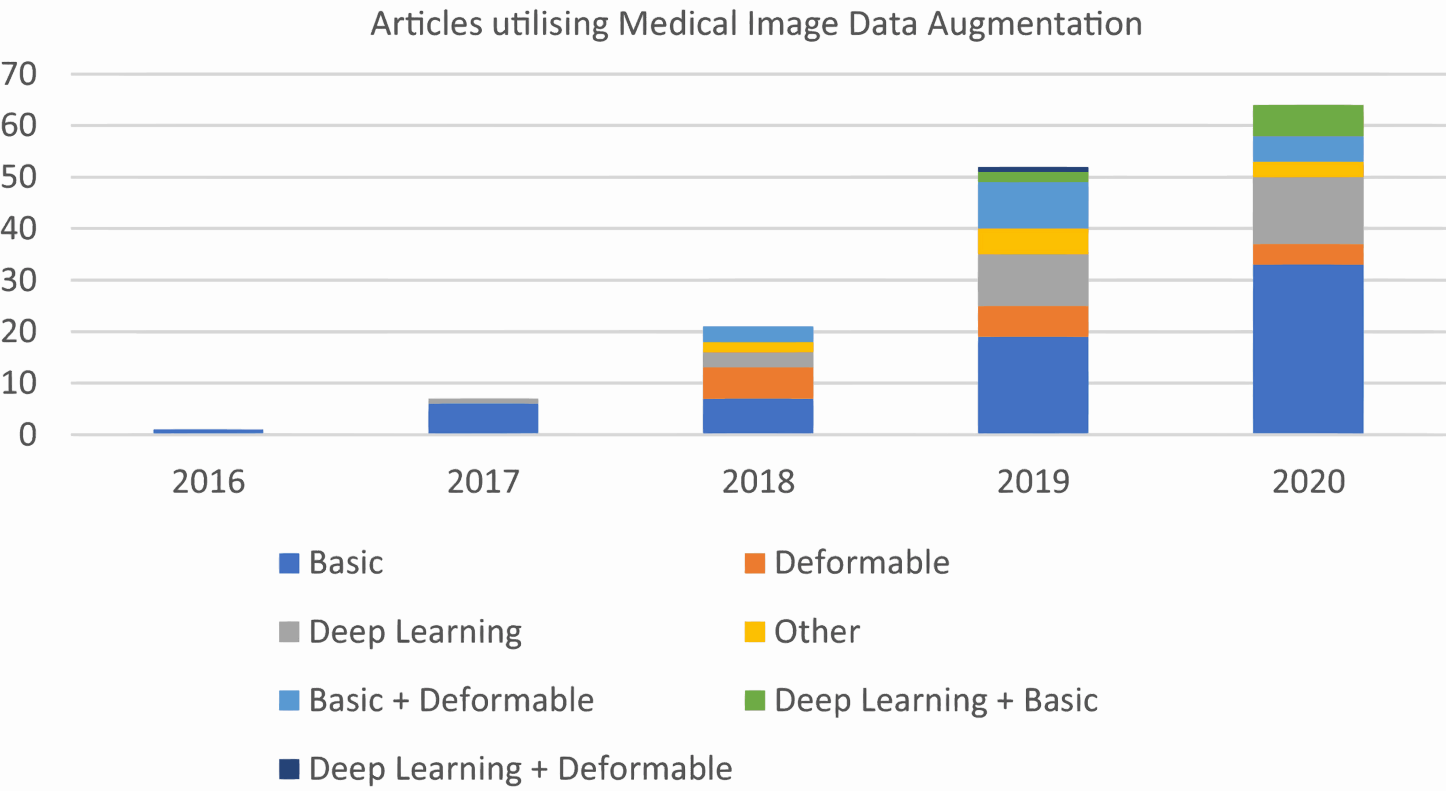}
  \caption{Results of Literature Review by Chlap\etal \cite{chlap2021review}.}
  \label{fig:literature_review}
\end{figure}

\begin{table}[htb]
\centering
\caption{Layer Segmentation: MSHC-Dataset - Full Training set}
\label{ms_5fold_nocorr_sall}
\resizebox{\linewidth}{!}{%
\begin{tabular}{l|lllllllll!{\vrule width \lightrulewidth}l}
\toprule
\textbf{Augmentation}                         & \textbf{ILM}                                  & \textbf{RNFL-GCL}                             & \textbf{IPL-INL}                              & \textbf{INL-OPL}                              & \textbf{OPL-ONL}                              & \textbf{ELM}                              & \textbf{IS-OS}                                  & \textbf{OS-RPE}                                & \textbf{BM}                                    & \textbf{All}  \\ 
\midrule
No Augmentation                                      & $0.76\pm0.46$                                      & $1.29\pm1.00^{\dagger}$                                      & $1.28\pm1.78$                                      & $1.19\pm1.13^{\dagger}$                                      & $1.08\pm0.85$                                      & $\mathbf{0.81\pm0.56}$                                      & $0.64\pm0.49^{\dagger}$                                      & $1.14\pm0.88$                                      & $1.05\pm1.11$                                      & $1.05\pm1.17^{\dagger}$         \\
{\cellcolor[rgb]{0.882,0.882,0.882}}Random Affine  & {\cellcolor[rgb]{0.882,0.882,0.882}}$0.79\pm0.57^{*}$  & {\cellcolor[rgb]{0.882,0.882,0.882}}$1.30\pm1.29$  & {\cellcolor[rgb]{0.882,0.882,0.882}}$1.23\pm1.75$  & {\cellcolor[rgb]{0.882,0.882,0.882}}$1.11\pm0.89^{\dagger}$  & {\cellcolor[rgb]{0.882,0.882,0.882}}$1.11\pm0.86^{\dagger}$  & {\cellcolor[rgb]{0.882,0.882,0.882}}$0.84\pm0.65$  & {\cellcolor[rgb]{0.882,0.882,0.882}}$0.64\pm0.55$  & {\cellcolor[rgb]{0.882,0.882,0.882}}$1.16\pm0.91^{\dagger}$  & {\cellcolor[rgb]{0.882,0.882,0.882}}$1.33\pm0.83^{*\dagger}$  & {\cellcolor[rgb]{0.882,0.882,0.882}}$1.08\pm1.17^{\dagger}$         \\
\end{tabular}
}
\end{table}
\vspace{-3em}
\begin{table}[htb]
\centering
\caption{Layer Segmentation: MSHC-Dataset - Limited Training set (4 samples with 7 slices)}
\label{appendix_da_ms_5fold_nocorr_s4}
\resizebox{\linewidth}{!}{%
\begin{tabular}{l|lllllllll!{\vrule width \lightrulewidth}l}
\toprule
\textbf{Augmentation}                         & \textbf{ILM}                                  & \textbf{RNFL-GCL}                             & \textbf{IPL-INL}                              & \textbf{INL-OPL}                              & \textbf{OPL-ONL}                              & \textbf{ELM}                              & \textbf{IS-OS}                                  & \textbf{OS-RPE}                                & \textbf{BM}                                    & \textbf{All}  \\ 
\midrule
No Augmentation                                      & $2.96\pm5.53$                                      & $\mathbf{3.13\pm3.93}$                                      & $3.94\pm6.23^{\dagger}$                                      & $3.15\pm5.36$                                      & $3.46\pm4.45^{\dagger}$                                      & $2.57\pm4.14^{\dagger}$                                      & $3.45\pm5.66^{\dagger}$                                      & $4.40\pm8.14^{\dagger}$                                      & $5.04\pm9.76$                                      & $3.64\pm6.77^{\dagger}$         \\
{\cellcolor[rgb]{0.882,0.882,0.882}}Random Affine  & {\cellcolor[rgb]{0.882,0.882,0.882}}$1.58\pm2.20$  & {\cellcolor[rgb]{0.882,0.882,0.882}}$3.13\pm3.01$  & {\cellcolor[rgb]{0.882,0.882,0.882}}$\mathbf{2.69\pm3.50}^{*}$  & {\cellcolor[rgb]{0.882,0.882,0.882}}$\mathbf{2.23\pm2.48}$  & {\cellcolor[rgb]{0.882,0.882,0.882}}$\mathbf{2.05\pm2.58}^{*}$  & {\cellcolor[rgb]{0.882,0.882,0.882}}$\mathbf{1.87\pm2.87}^{*}$  & {\cellcolor[rgb]{0.882,0.882,0.882}}$\mathbf{1.84\pm3.28}^{*}$  & {\cellcolor[rgb]{0.882,0.882,0.882}}$\mathbf{2.08\pm2.95}^{*}$  & {\cellcolor[rgb]{0.882,0.882,0.882}}$\mathbf{2.98\pm4.52}$  & {\cellcolor[rgb]{0.882,0.882,0.882}}$\mathbf{2.33\pm3.30}^{*}$         \\
\end{tabular}
}
\end{table}
\vspace{-3em}
\begin{table}[htb]
\centering
\caption{Layer Segmentation: MSHC-Dataset - Limited Training set (2 samples with 7 slices)}
\label{ms_5fold_nocorr_s2}
\resizebox{\linewidth}{!}{%
\begin{tabular}{l|lllllllll!{\vrule width \lightrulewidth}l}
\toprule
\textbf{Augmentation}                         & \textbf{ILM}                                  & \textbf{RNFL-GCL}                             & \textbf{IPL-INL}                              & \textbf{INL-OPL}                              & \textbf{OPL-ONL}                              & \textbf{ELM}                              & \textbf{IS-OS}                                  & \textbf{OS-RPE}                                & \textbf{BM}                                    & \textbf{All}  \\ 
\midrule
No Augmentation                                      & $2.68\pm4.09$                                      & $5.34\pm5.17^{\dagger}$                                      & $5.65\pm7.13^{\dagger}$                                      & $5.61\pm7.48^{\dagger}$                                      & $4.56\pm4.73^{\dagger}$                                      & $6.85\pm8.39^{\dagger}$                                      & $6.37\pm8.48^{\dagger}$                                      & $7.30\pm9.81^{\dagger}$                                      & $9.05\pm13.03^{\dagger}$                                      & $6.17\pm9.00^{\dagger}$         \\
{\cellcolor[rgb]{0.882,0.882,0.882}}Random Affine  & {\cellcolor[rgb]{0.882,0.882,0.882}}$2.98\pm6.35$  & {\cellcolor[rgb]{0.882,0.882,0.882}}$\mathbf{4.36\pm6.29}^{*}$  & {\cellcolor[rgb]{0.882,0.882,0.882}}$\mathbf{3.77\pm4.44}^{*}$  & {\cellcolor[rgb]{0.882,0.882,0.882}}$\mathbf{2.76\pm3.07}^{*}$  & {\cellcolor[rgb]{0.882,0.882,0.882}}$\mathbf{2.86\pm2.97}^{*}$  & {\cellcolor[rgb]{0.882,0.882,0.882}}$\mathbf{2.51\pm3.66}^{*}$  & {\cellcolor[rgb]{0.882,0.882,0.882}}$\mathbf{2.30\pm3.57}^{*}$  & {\cellcolor[rgb]{0.882,0.882,0.882}}$\mathbf{2.74\pm3.64}^{*}$  & {\cellcolor[rgb]{0.882,0.882,0.882}}$\mathbf{2.66\pm3.85}^{*}$  & {\cellcolor[rgb]{0.882,0.882,0.882}}$\mathbf{3.06\pm4.77}^{*}$         \\
\end{tabular}
}
\end{table}
\vspace{-3em}
\begin{table}[htb]
\centering
\caption{Layer Segmentation: Multibiomarker-Dataset - Full Training set}
\label{multibiomarker_5fold_nocorr_sall}
\resizebox{\linewidth}{!}{%
\begin{tabular}{l|lllllllllll!{\vrule width \lightrulewidth}l} 
\toprule
\textbf{Augmentation}                         & \textbf{ILM}                                  & \textbf{RNFL-GCL}                             & \textbf{GCL-IPL}                              & \textbf{IPL-INL}                              & \textbf{INL-OPL}                              & \textbf{OPL-HFL}                              & \textbf{ELM}                                  & \textbf{BMEIS}                                & \textbf{IB-RPE}                               & \textbf{OB-RPE}                                & \textbf{BM}                                    & \textbf{All}  \\ 
\midrule
No Augmentation                                      & $2.09\pm4.37$                                      & $1.73\pm1.96$                                      & $2.31\pm2.92$                                      & $2.21\pm3.21$                                      & $2.69\pm3.44^{\dagger}$                                      & $2.84\pm3.32^{\dagger}$                                      & $2.25\pm2.25$                                      & $1.74\pm1.74$                                      & $2.42\pm4.13$                                      & $2.23\pm3.84$                                      & $3.35\pm5.35^{\dagger}$                                      & $2.39\pm3.78^{\dagger}$         \\
\end{tabular}
}
\end{table}
\vspace{-3em}
\begin{table}[htb]
\centering
\caption{Layer Segmentation: Multibiomarker-Dataset - Limited Training set (4 samples with 7 slices)}
\label{appendix_da_multibiomarker_5fold_nocorr_s4}
\resizebox{\linewidth}{!}{%
\begin{tabular}{l|lllllllllll!{\vrule width \lightrulewidth}l} 
\toprule
\textbf{Augmentation}                         & \textbf{ILM}                                  & \textbf{RNFL-GCL}                             & \textbf{GCL-IPL}                              & \textbf{IPL-INL}                              & \textbf{INL-OPL}                              & \textbf{OPL-HFL}                              & \textbf{ELM}                                  & \textbf{BMEIS}                                & \textbf{IB-RPE}                               & \textbf{OB-RPE}                                & \textbf{BM}                                    & \textbf{All}  \\ 
\midrule
No Augmentation                                      & $5.21\pm9.95^{\dagger}$                                      & $7.47\pm11.48^{\dagger}$                                      & $7.39\pm9.55^{\dagger}$                                      & $8.02\pm9.65^{\dagger}$                                      & $7.76\pm8.40^{\dagger}$                                      & $8.00\pm8.67^{\dagger}$                                      & $5.70\pm6.11^{\dagger}$                                      & $6.08\pm8.66^{\dagger}$                                      & $5.02\pm6.52$                                      & $5.10\pm6.48$                                      & $13.70\pm17.12^{\dagger}$                                      & $7.59\pm11.19^{\dagger}$         \\
\end{tabular}
}
\end{table}
\vspace{-3em}
\begin{table}[htb]
\centering
\caption{Layer Segmentation: Multibiomarker-Dataset - Limited Training set (2 samples with 7 slices)}
\label{multibiomarker_5fold_nocorr_s2}
\resizebox{\linewidth}{!}{%
\begin{tabular}{l|lllllllllll!{\vrule width \lightrulewidth}l} 
\toprule
\textbf{Augmentation}                         & \textbf{ILM}                                  & \textbf{RNFL-GCL}                             & \textbf{GCL-IPL}                              & \textbf{IPL-INL}                              & \textbf{INL-OPL}                              & \textbf{OPL-HFL}                              & \textbf{ELM}                                  & \textbf{BMEIS}                                & \textbf{IB-RPE}                               & \textbf{OB-RPE}                                & \textbf{BM}                                    & \textbf{All}  \\ 
\midrule
No Augmentation                                      & $10.70\pm27.63^{\dagger}$                                      & $11.26\pm22.79^{\dagger}$                                      & $15.75\pm25.01^{\dagger}$                                      & $15.39\pm25.40^{\dagger}$                                      & $14.20\pm21.09^{\dagger}$                                      & $15.10\pm22.94^{\dagger}$                                      & $13.23\pm19.99^{\dagger}$                                      & $14.86\pm21.87^{\dagger}$                                      & $13.98\pm20.62^{\dagger}$                                      & $13.60\pm22.12^{\dagger}$                                      & $24.68\pm33.26^{\dagger}$                                      & $15.20\pm25.07^{\dagger}$         \\
\end{tabular}
}
\end{table}
\vspace{-3em}
\begin{figure}[htb]
  \centering
  \includegraphics[width=1.0\textwidth]{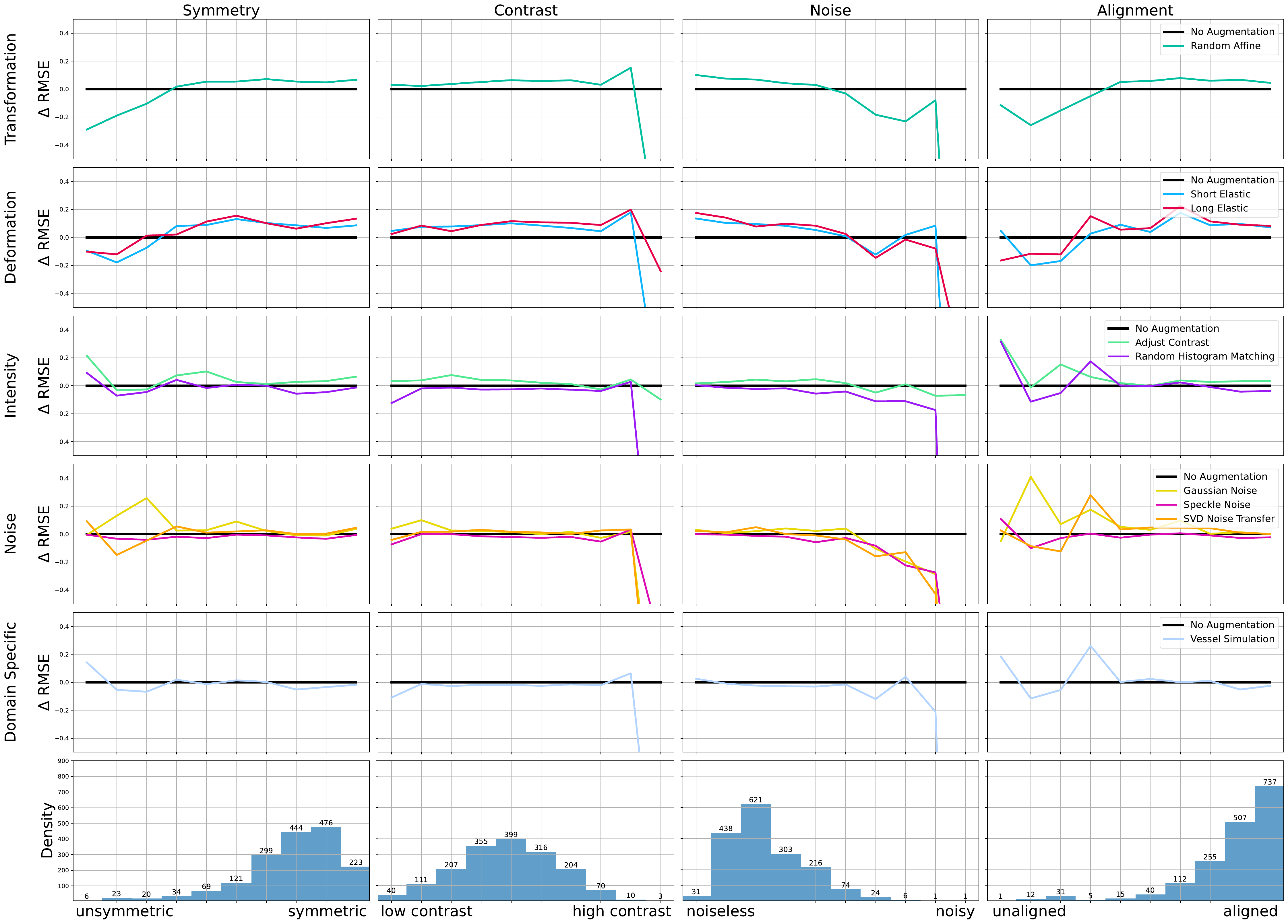}
   \caption{MSHC-Dataset: RMSE distance with respect to OCT Scan metrics.}
  \label{fig:ms_full_RMSE_Metrics}
\end{figure}

\begin{figure}[htb]
  \centering
  \includegraphics[width=1.0\textwidth]{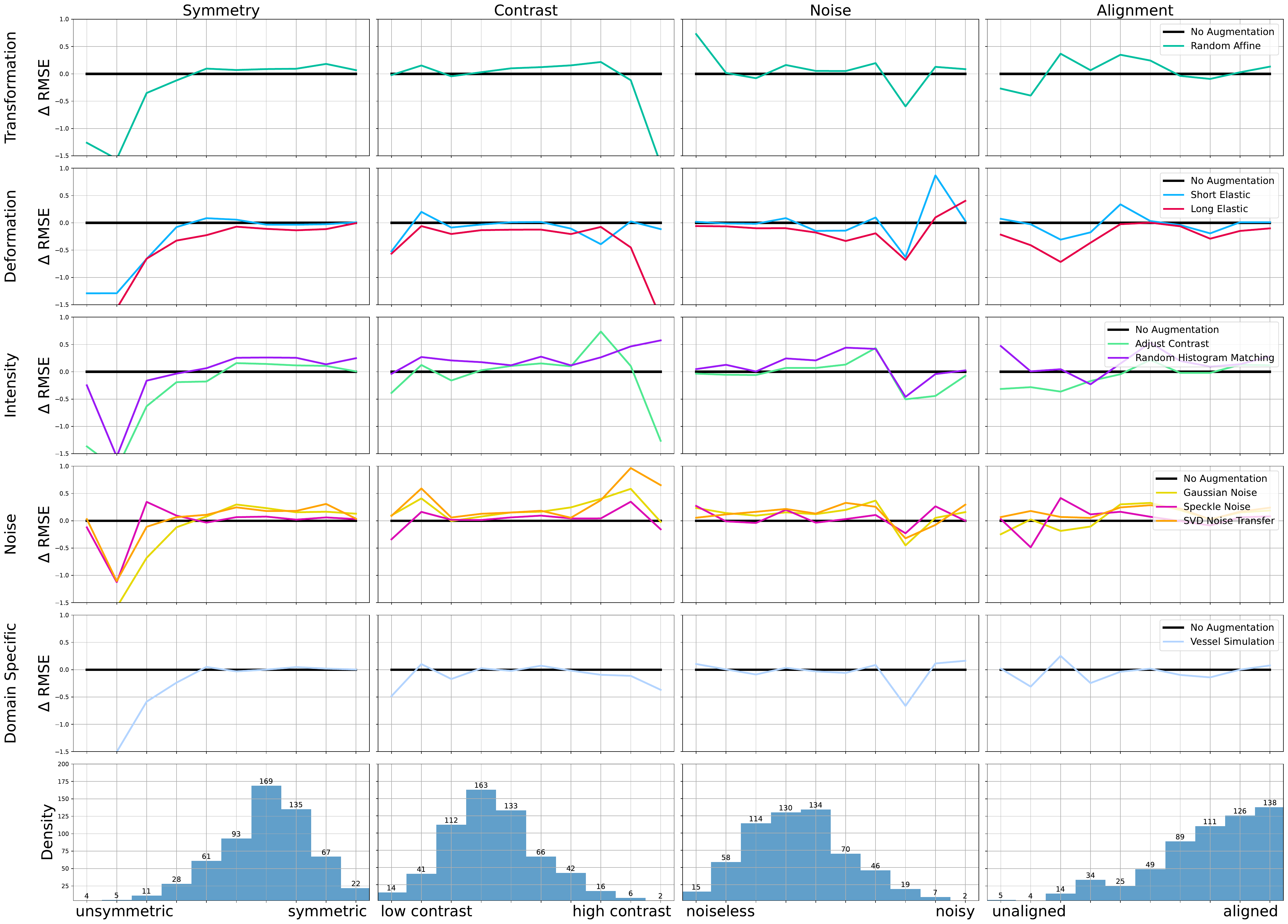}
   \caption{Multibiomarker-Dataset: RMSE distance with respect to OCT Scan metrics.}
  \label{fig:multibiomarker_full_RMSE_Metrics}
\end{figure}

\clearpage
\bibliographystyle{splncs04}
\bibliography{Paper-0022}